\documentclass[aps,prd,preprint,superscriptaddress,groupedaddress,showpacs,amsmath,amssymb]{revtex4}
\usepackage[dvips]{graphicx}
\usepackage{amsmath,amssymb,times}

\newcommand{\bequ}{\begin{equation}}
\newcommand{\eequ}{\end{equation}}
\newcommand{\bea}{\begin{eqnarray}}
\newcommand{\eea}{\end{eqnarray}}

\DeclareSymbolFont{boldletters}{OML}{cmm} {b}{it}
\DeclareSymbolFontAlphabet{\mathbit}{boldletters}
\DeclareMathSymbol{\alpha}{\mathalpha}{letters}{"0B}
\DeclareMathSymbol{\beta}{\mathalpha}{letters}{"0C}
\DeclareMathSymbol{\gamma}{\mathalpha}{letters}{"0D}
\DeclareMathSymbol{\delta}{\mathalpha}{letters}{"0E}
\DeclareMathSymbol{\epsilon}{\mathalpha}{letters}{"0F}
\DeclareMathSymbol{\zeta}{\mathalpha}{letters}{"10}
\DeclareMathSymbol{\eta}{\mathalpha}{letters}{"11}
\DeclareMathSymbol{\theta}{\mathalpha}{letters}{"12}
\DeclareMathSymbol{\iota}{\mathalpha}{letters}{"13}
\DeclareMathSymbol{\kappa}{\mathalpha}{letters}{"14}
\DeclareMathSymbol{\lambda}{\mathalpha}{letters}{"15}
\DeclareMathSymbol{\mu}{\mathalpha}{letters}{"16}
\DeclareMathSymbol{\nu}{\mathalpha}{letters}{"17}
\DeclareMathSymbol{\xi}{\mathalpha}{letters}{"18}
\DeclareMathSymbol{\pi}{\mathalpha}{letters}{"19}
\DeclareMathSymbol{\rho}{\mathalpha}{letters}{"1A}
\DeclareMathSymbol{\sigma}{\mathalpha}{letters}{"1B}
\DeclareMathSymbol{\tau}{\mathalpha}{letters}{"1C}
\DeclareMathSymbol{\upsilon}{\mathalpha}{letters}{"1D}
\DeclareMathSymbol{\phi}{\mathalpha}{letters}{"1E}
\DeclareMathSymbol{\chi}{\mathalpha}{letters}{"1F}
\DeclareMathSymbol{\psi}{\mathalpha}{letters}{"20}
\DeclareMathSymbol{\omega}{\mathalpha}{letters}{"21}
\DeclareMathSymbol{\varepsilon}{\mathalpha}{letters}{"22}
\DeclareMathSymbol{\vartheta}{\mathalpha}{letters}{"23}
\DeclareMathSymbol{\varpi}{\mathalpha}{letters}{"24}
\DeclareMathSymbol{\varrho}{\mathalpha}{letters}{"25}
\DeclareMathSymbol{\varsigma}{\mathalpha}{letters}{"26}
\DeclareMathSymbol{\varphi}{\mathalpha}{letters}{"27}
\DeclareMathSymbol{\Gamma}{\mathalpha}{letters}{"00}
\DeclareMathSymbol{\Delta}{\mathalpha}{letters}{"01}
\DeclareMathSymbol{\Theta}{\mathalpha}{letters}{"02}
\DeclareMathSymbol{\Lambda}{\mathalpha}{letters}{"03}
\DeclareMathSymbol{\Xi}{\mathalpha}{letters}{"04}
\DeclareMathSymbol{\Pi}{\mathalpha}{letters}{"05}
\DeclareMathSymbol{\Sigma}{\mathalpha}{letters}{"06}
\DeclareMathSymbol{\Upsilon}{\mathalpha}{letters}{"07}
\DeclareMathSymbol{\Phi}{\mathalpha}{letters}{"08}
\DeclareMathSymbol{\Psi}{\mathalpha}{letters}{"09}
\DeclareMathSymbol{\Omega}{\mathalpha}{letters}{"0A}

\begin{document}
\preprint{SAGA-HE-265, RIKEN-QHP-53}
\title{ 
The quarkynic phase and the ${Z}_{N_c}$ symmetry}

\author{Yuji Sakai}
\email[]{ysakai@riken.ac.jp}
\affiliation{Quantum Hadron Physics Laboratory RIKEN Nishina Center, 
	Saitama 351-0198, Japan}

\author{Hiroaki Kouno}
\email[]{kounoh@cc.saga-u.ac.jp}
\affiliation{Department of Physics, Saga University,
             Saga 840-8502, Japan}

\author{Takahiro Sasaki}
\email[]{sasaki@phys.kyushu-u.ac.jp}
\affiliation{Department of Physics, Graduate School of Sciences, Kyushu University,
             Fukuoka 812-8581, Japan}

\author{Masanobu Yahiro}
\email[]{yahiro@phys.kyushu-u.ac.jp}
\affiliation{Department of Physics, Graduate School of Sciences, Kyushu University,
             Fukuoka 812-8581, Japan}

\date{\today}

\begin{abstract}
We investigate the interplay between the $\mathbb{Z}_{N_c}$ symmetry and 
the emergence of the quarkyonic phase, adding the flavor-dependent 
complex chemical potentials 
$\mu_f=\mu+i T \theta_f$ with $(\theta_f)=(0,\theta,-\theta)$ to 
the Polyakov-loop extended Nambu--Jona-Lasinio (PNJL) model. 
When $\theta=0$, the PNJL model with the $\mu_f$ agrees 
with the standard PNJL model 
with the real chemical potential $\mu$.  
When $\theta=2\pi/3$, meanwhile, 
the PNJL model with the $\mu_f$ has the $\mathbb{Z}_{N_c}$ symmetry exactly 
for any real $\mu$, so that the quarkyonic phase exists 
at small $T$ and large $\mu$. 
Once $\theta$ varies from $2\pi/3$, the quarkyonic phase exists only 
on a line of $T=0$ and $\mu$ larger than the dynamical quark mass, 
and the region 
at small $T$ and large $\mu$ is dominated by 
the quarkyonic-like phase in which the Polyakov loop is small but finite. 
\end{abstract}

\pacs{11.30.Rd, 12.40.-y}
\maketitle

%%%%%%%%%%%%%%%%%%%%%%%%%%%%%%%%%%%%%%%%%%%%%%%%%%%%%%%%%%%%%%%%%%%%%%%%%%%
%%%%%  Introduction 
%%%%%%%%%%%%%%%%%%%%%%%%%%%%%%%%%%%%%%%%%%%%%%%%%%%%%%%%%%%%%%%%%%%%%%%%%%%
%\section{Introduction}
 
Understanding of the confinement mechanism is 
one of the most important subjects in hadron physics. 
Lattice QCD (LQCD) shows numerically that QCD is in the confinement 
and chiral symmetry breaking phase at low temperature ($T$) and 
in the deconfinement and chiral symmetry restoration phase at high $T$. 
In the limit of infinite current quark mass, 
the Polyakov-loop is an exact order parameter for the deconfinement transition, since the $\mathbb{Z}_{N_c}$ symmetry is exact there. 
The chiral condensate is, meanwhile, an exact order parameter 
for the chiral restoration in the limit of zero current quark mass. 
In the real world where $u$ and $d$ quarks have small current masses, 
the chiral condensate is considered to be a good order parameter 
for the chiral restoration, but there is no guarantee 
that the Polyakov-loop is a good order parameter for 
the deconfinement transition.

In the previous paper~\cite{Kouno_TBC}, 
we have proposed a QCD-like theory 
with the $\mathbb{Z}_{N_c}$ symmetry. 
Let us start with the SU($N_c$) gauge theory with $N_f$ 
degenerate flavors to construct the QCD-like theory. 
The partition function $Z$ of the SU($N_c$) gauge theory
is obtained in Euclidean space-time by 
%%%%%%%%%%%%%%%%
\bea
Z=\int Dq D\bar{q} DA \exp[-S_0] 
\label{QCD-Z}
\eea
%%%%%%%%%%%%%%%%
with the action
\bea
S_0=\int d^4x [\sum_{f}\bar{q}_f(\gamma_\nu D_\nu +m_f)q_f
+{1\over{4g^2}}F_{\mu\nu}^2], 
\label{QCD-S}
\eea
%%%%%%%%%%%%%
where $q_f$ is the quark field with flavor $f$ and current quark mass $m_f$, 
$D_\nu =\partial_\nu-iA_\nu$ is the covariant derivative 
with the gauge field $A_\nu$, $g$ is the gauge coupling and 
$F_{\mu\nu}=\partial_\mu A_\nu -\partial_\nu A_\mu -i[A_\mu ,A_\nu ]$. 
The temporal boundary condition for quark is 
%%%%%%%%%%%%%%%%
\bea
q_f(x, \beta=1/T)=-q_f(x, 0). 
\label{period-QCD}
\eea
%%%%%%%%%%%%%%%%
The fermion boundary condition is changed by 
the $\mathbb{Z}_{N_c}$ transformation as~\cite{RW,Sakai}
%%%%%%%%%%%%%%%%
\bea
q_f(x, \beta)=-\exp{(-i 2\pi k/{N_c})}q_f(x, 0)  
\label{period-QCD-Z}
\eea
%%%%%%%%%%%%%%%%
for integer $k$, while the action $S_0$ keeps the form of \eqref{QCD-S} 
in virtue of the fact that the $\mathbb{Z}_{N_c}$ symmetry is 
the center symmetry of the gauge symmetry~\cite{RW}. 
The $\mathbb{Z}_{N_c}$ symmetry thus breaks down 
through the fermion boundary condition \eqref{period-QCD} in QCD. 

Now we consider the SU($N$) gauge theory with $N$ degenerate flavors, i.e. 
$N=N_c=N_f$, and assume 
the twisted boundary condition (TBC) in the temporal direction~\cite{Kouno_TBC}:
%%%%%%%%%%%%%%%%
\begin{eqnarray}
q_f(x, \beta )=-\exp(-i\theta_f)q_f(x,0) 
\label{period}
\end{eqnarray}
%%%%%%%%%%%%%%%%
with the twisted angles
%%%%%%%%%%%%%%%%
\begin{eqnarray}
\theta_f=2\pi (f-1)/N+\theta_1
\label{twisted-angle}
\end{eqnarray}
%%%%%%%%%%%%%%%%
for flavors $f$ labeled by integers from $1$ to $N$, where $\theta_1$ 
is an arbitrary real number in a range of $0 \le \theta_1 < 2\pi$. 
The action $S_0$ with the TBC is a QCD-like theory proposed 
in Ref.~\cite{Kouno_TBC}. In fact, the QCD-like theory has 
the $\mathbb{Z}_{N_c}$ symmetry, since 
$f$ is changed into $f+k$ by the $\mathbb{Z}_{N}$ transformation but 
$f+k$ can be relabeled by $f$. 
In the QCD-like theory, the Polyakov loop becomes an exact order parameter 
of the deconfinement transition. The QCD-like theory then becomes 
a quite useful theory to understand the confinement mechanism.

When the fermion field $q_f$ is transformed by 
%%%%%%%%%%%%%%
\begin{eqnarray}
q_f \to \exp{(-i\theta_fT\tau)}q_f 
\label{transform_1}
\end{eqnarray}
%%%%%%%%%%%%%%%
with the twisted angle $\theta_f$ and the Euclidean time $\tau$, 
the action $S_0$ is changed into 
%%%%%%%%%%%%%%%%
\begin{eqnarray}
S(\theta_f)=\int d^4x [\sum_{f}\bar{q}_f(\gamma_\nu D_\nu -i\theta_fT\gamma_4+m_f)q_f
+{1\over{4g^2}}F_{\mu\nu}^2] 
\label{QCD1}
\end{eqnarray}
%%%%%%%%%%%%%
with the imaginary chemical potential $\mu_f=i T \theta_f$, while  
the TBC returns to the standard one \eqref{period-QCD}. 
The action $S_0$ with the TBC is thus identical with the action 
$S(\theta_f)$ with the standard one \eqref{period-QCD}. 
In the limit of $T=0$, the action $S(\theta_f)$ comes back to the QCD action 
$S_0$ with the standard boundary condition \eqref{period-QCD} kept. 
The QCD-like theory thus agrees with QCD at $T=0$ where 
the Polyakov loop $\Phi$ is zero. One can then expect  
that in the QCD-like theory $\Phi$ is zero up to some temperature $T_c$ and 
becomes finite above $T_c$, i.e, that the $\mathbb{Z}_{N_c}$ symmetry is 
exactly preserved below $T_c$ but spontaneously broken above $T_c$. 
Actually, this behavior is confirmed by imposing the TBC on 
the Polyakov-loop extended Nambu--Jona-Lasinio (PNJL) model\cite{Meisinger,Dumitru,Fukushima,Ratti,Rossner,Fukushima2,Abuki,Sakai,Matsumoto,Sakai_imiso,Sakai5,Hell,Gatto,Sasaki-T_Nf3,Sakai_hadron,McLerran_largeNc,Schaefer,BL,Kashiwa1}. 
The PNJL model with the TBC~\cite{Kouno_TBC} is referred to  as the TBC model in this paper. 
In the TBC model, the flavor symmetry is explicitly broken by 
the flavor-dependent TBC \eqref{period}, but the flavor-symmetry breaking 
is recovered at $T<T_c$. 
The TBC model is thus a model proper to understand the confinement mechanism.

A current topic related to the confinement is the quarkyonic phase~\cite{McLerran1,Fukushima2,Abuki,McLerran_largeNc,BL}. 
It is a confined (color-singlet) phase with finite quark-number density $n$, 
that is, a phase with $\Phi=0$ and $n \ne 0$. 
The $n$-generation induces the chiral restoration; in fact, 
the two phenomena occur almost simultaneously in the PNJL model 
\cite{Sakai_hadron}. This fact indicates that 
the quarkyonic phase can be regarded as a chirally-symmetric 
and confined phase. It was suggested 
in Refs.~\cite{Nakano,Nickel} 
that the chirally-broken phase is enlarged toward 
lager $\mu$ by the chiral density wave. 
In this paper, for simplicity, we ignore 
inhomogeneous condensates such as the chiral density wave. 
Effects of the inhomogeneous condensate on the quarkynic phase 
and the interplay between the effects and the $\mathbb{Z}_{N_c}$ symmetry 
are interesting as a future work. 
The concept of the quarkyonic phase was constructed in large $N_c$ QCD. 
In fact, the phase was first found at small $T$ and large 
real quark-number chemical potential $\mu$ in large $N_c$ QCD. 
Recently, the PNJL model showed that 
a quarkyonic-like phase with $\Phi < 0.5$ and $n \ne 0$ exists 
at small $T$ and large $\mu$ for the case of $N_c=3$\cite{McLerran_largeNc,BL}. 
This result may stem from the fact that 
the deconfinement transition is crossover for $N_c=3$. 
This suggests that 
the quarkyonic phase can survive even at $N_c=3$ in the QCD-like theory with 
the $\mathbb{Z}_{N_c}$ symmetry.

%%%%%%%%%%%%%%%
\begin{figure}[htbp]
\begin{center}
\includegraphics[width=0.3\textwidth]{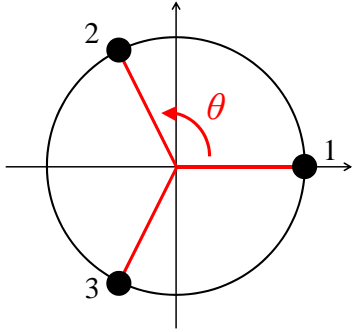}
\end{center}
\caption{
Location of $\exp[i\theta_f]$ in the complex plane; here, 
$(\theta_f)=(0,\theta,-\theta)$. 
}
\label{Sch}
\end{figure}
%%%%%%%%%%%%

In this paper, we consider the PNJL model of $N \equiv N_c=N_f=3$ with 
the flavor-independent real chemical potential $\mu$ and 
the flavor-dependent quark 
boundary condition \eqref{period} with 
\bea
(\theta_f)=(0,\theta,-\theta) 
\label{NBC}
\eea
instead of \eqref{twisted-angle};
see Fig. \ref{Sch} for the boundary condition.
The present system is the same as that 
with the standard boundary condition \eqref{period-QCD} and 
the flavor-dependent complex chemical potentials 
$\mu_f = \mu+i T \theta_f$ with \eqref{NBC}. 
The present model with the $\mu_f$ is reduced to the standard PNJL model 
with the flavor-independent  real chemical potential $\mu$ 
when $\theta=0$ and 
to the TBC model with the $\mathbb{Z}_{N_c}$ symmetry 
when $\theta=2\pi/3$. 
Varying $\theta$, one can see how the phase diagram is 
changed between the exact color-confinement 
in the TBC model and the approximate one in the standard PNJL model. 
The aim of this paper is to see this behavior. 
Our particular interest is the location of the quarkyonic and 
the quarkyonic-like phase in the $\mu$-$T$ plane.

In general, there is no guarantee that the QCD partition function 
with complex chemical potential is real. 
It is, however, possible to prove that  
the QCD partition function $Z_0(\mu_f)$ with $\mu_f = \mu+i T \theta_f$ 
is real. 
The fermion  determinant $\det{\cal M}(\mu_f)$ satisfies the relation
%%%%%%%%%%%%%%%%%%%%%%%%%%
\begin{eqnarray}
\left[\det{\cal M}(\mu_f)\right]^*&=&
\det{\cal M}(-\mu_f^*)\notag\\
&=&\prod_f\det[D-(\mu-i\theta_fT)\gamma_4+m_f]
\notag\\
&=&\prod_f\det[D-(\mu+i\theta_fT)\gamma_4+m_f]
=\det{\cal M}(-\mu_f),
\label{fermion-det}
\end{eqnarray} 
%%%%%%%%%%%%%%%%%%%%%%%%%%
where the third equality is obtained by the relabeling of the $f$. 
The present system thus has the sign problem, but the partition function 
is real, since  
\bea
Z_0(\mu_f)^*=Z_0(-\mu_f)=Z_0(\mu_f), 
\label{symm}
\eea
where 
the first equality is obtained by \eqref{fermion-det} and the second one 
by the charge conjugation. 
Also in the PNJL model with the $\mu_f$, the partition funciton 
is real, as shown later.

%%%%%%%%%%%%%%%%%%%%%%%%%%
%\section{Formalism}
%\label{Formalism}
%%%%%%%%%%%%%%%%%%%%%%%%%%

The three-flavor PNJL Lagrangian is defined in Euclidian space-time as
%%%%%%%%%%%%%%%%%%%%%%%%%%
\begin{align}
{\cal L}&={\bar q}(\gamma_\nu D_\nu+{\hat m}-\mu\gamma_4)q  
-G_{\rm S}
\sum_{a=0}^{8}[({\bar q}\lambda_a q)^2+({\bar q}i\gamma_5\lambda_a q )^2] 
\nonumber\\
&+G_{\rm D}\left[\det_{ij}{\bar q}_i(1+\gamma_5)q_j+{\rm h.c.}\right]
+{\cal U}(\Phi [A],\Phi^*[A],T),
\label{L_nc3}
\end{align} 
%%%%%%%%%%%%%%%%%%%%%%%%%%
where $D_\nu=\partial_\nu -i\delta_{\nu4}A_4$, $\lambda_a$ is the Gell-Mann 
matrices and ${\hat m}={\rm diag}(m_1,m_2,m_3)$. 
$G_{\rm S}$ and $G_{\rm D}$ are coupling constants 
of the scalar-type four-quark and the Kobayashi-Maskawa-'t Hooft (KMT) 
interaction~\cite{KMK,tHooft}, respectively. 
The KMT interaction breaks the $U_\mathrm{A} (1)$ symmetry explicitly. 
The Polyakov-loop $\Phi$ and its conjugate $\Phi^*$ are defined by
%%%%%%%%%%%%%%%
% gauge field %
%%%%%%%%%%%%%%% 
\begin{align}
\Phi &= {1\over{3}}{\rm tr}_c(L),\quad
\Phi^* ={1\over{3}}{\rm tr}_c({\bar L}),
\label{Polyakov_nc3}
\end{align}
%%%%%%%%%%%
with $L=\exp(i A_4/T)$ in the Polyakov gauge.
We take the Polyakov potential of Ref.~\cite{Rossner}:
%%%%%%%%%%%%%%%%
\begin{align}
&{\cal U} = T^4 \Bigl[-\frac{a(T)}{2} {\Phi}^*\Phi
+ b(T)\ln(1 - 6{\Phi\Phi^*}  + 4(\Phi^3+{\Phi^*}^3)
- 3(\Phi\Phi^*)^2 )\Bigr] ,
\label{eq:E13}\\
&a(T)   = a_0 + a_1\Bigl(\frac{T_0}{T}\Bigr)
                 + a_2\Bigl(\frac{T_0}{T}\Bigr)^2,~~~~
b(T)=b_3\Bigl(\frac{T_0}{T}\Bigr)^3 .
            \label{eq:E14}
\end{align}
%%%%%%%%%%%%%%
Parameters of $\mathcal{U}$ are fitted to LQCD data 
at finite $T$ in the pure gauge limit. 
The parameters except $T_0$ are summarized in Table \ref{table-para}.  
The Polyakov potential yields the first-order deconfinement phase transition 
at $T=T_0$ in the pure gauge theory~\cite{Boyd,Kaczmarek}. 
The original value of $T_0$ is $270$ MeV determined from the pure gauge 
LQCD data, but the PNJL model with this value yields a larger 
value of the pseudocritical temperature $T_\mathrm{c}$ 
at zero chemical potential than $T_c\approx 160$~MeV predicted 
by full LQCD \cite{Borsanyi,Soeldner,Kanaya}. 
We then rescale $T_0$ to 195~MeV so as to reproduce 
$T_c=160$~MeV~\cite{Sasaki-T_Nf3}. 
%%%%%%%%%%%%%%%%
%%% Table I
%%%%%%%%%%%%%%%%
\begin{table}[h]
\begin{center}
\begin{tabular}{llllll}
\hline \hline
~~~~~$a_0$~~~~~&~~~~~$a_1$~~~~~&~~~~~$a_2$~~~~~&~~~~~$b_3$~~~~~
\\
\hline
~~~~3.51 &~~~~-2.47 &~~~~15.2 &~~~~-1.75\\
\hline \hline
\end{tabular}
\caption{
Summary of the parameter set in the Polyakov-potential sector 
determined in Ref.~\cite{Rossner}. 
All parameters are dimensionless. 
}
\label{table-para}
\end{center}
\end{table}
%%%%%%%%%%%

Now we consider the flavor-dependent complex chemical potential 
$\mu_f=\mu+i\theta_fT$. 
The thermodynamic potential 
(per volume) is obtained by the mean-field approximation as~\cite{Matsumoto}  
%%%%%%%%%%%%%
\begin{align}
\Omega
=\Omega_{\rm Q}(\sigma_f,\Phi,T,\mu_f)+U_{\rm M}(\sigma_f) +{\cal U}(\Phi,T) 
\label{PNJL-Omega_original}
\end{align}
with 
\begin{align}
\Omega_{\rm Q}&= -2\sum_{f=1}^3\int\frac{d^3{\bf p}}{(2\pi)^3}
{\rm tr_c}\Bigl[E_f+\frac{1}{\beta}\ln\left(1+Le^{-\beta E^-_f}\right)
+\frac{1}{\beta}\ln\left(1+Le^{-\beta E^+_f}\right)\Bigl], 
\label{PNJL-Omega_original-1}
\end{align}
%%%%%%%%%%%
where $\sigma_{f}=\langle{\bar q}_fq_f\rangle$, $E^{\pm}_f=E_f\pm\mu_f$
and $E_f=\sqrt{{\bf p}^2+{M_{f}}^2}$. 
Here the three-dimensional cutoff is taken for the momentum integration 
in the vacuum term~\cite{Matsumoto}. 
Obviously, $\Omega$ is real. 
The dynamical quark masses $M_{f}$ and the mesonic potential $U_{\rm M}$ 
are defined by 
%%%%%%%%%%%%%
\begin{align}
M_{f}&=m_{f}-4G_{\rm S}\sigma_{f}+
2G_{\rm D}|\epsilon_{fgh}|\sigma_{g}\sigma_{h}, \\
U_{\rm M}&=\sum_{f}  2 G_{\rm S} \sigma_{f}^2 
-4 G_{\rm D} \sigma_{1}\sigma_{2}\sigma_{3}, 
\end{align}
%%%%%%%%%%%
where $\epsilon_{fgh}$ is the antisymmetric symbol.

The PNJL model has six parameters, 
($G_{\rm S}$, $G_{\rm D}$, $m_1$, $m_2$, $m_3$, $\Lambda$). 
A typical set of the parameters is obtained in Ref.~\cite{Rehberg} 
for the 2+1 flavor system with $m_1=m_2\equiv m_l < m_3$. 
The parameter set is fitted to empirical values of 
$\eta'$-meson mass and $\pi$-meson mass and 
$\pi$-meson decay constant at vacuum. 
In the present paper, we set $m_0 \equiv m_l=m_3$ in the parameter set 
of Ref.~\cite{Rehberg}. 
The parameter set thus determined is shown in Table~\ref{Table_NJL}. 
%%%%%%%%%%%%%%%%
%%% Table II
%%%%%%%%%%%%%%%%
\begin{table}[h]
\begin{center}
\begin{tabular}{llllll}
\hline
~~$m_0(\rm MeV)$~~&~~$\Lambda(\rm MeV)$~~~&~~$G_{\rm S} \Lambda^2$
~~&~~$G_{\rm D} \Lambda^5$~~
\\
\hline
~~~~~~~~5.5 &~~~~~~~~602.3 &~~~~~~1.835 &~~~~~~12.36 &~~~~\\
\hline
\end{tabular}
\caption{
Summary of the parameter set in the NJL sector. 
All the parameters except $m_0$ are the same as in Ref.~\cite{Rehberg}. 
\label{Table_NJL}
}
\end{center}
\end{table}
%%%%%%%%%%%

Taking the color summation in \eqref{PNJL-Omega_original} leads to 
%%%%%%%%%%%%%
\begin{align}
\Omega
=-2\sum_{f}\int\frac{d^3{\bf p}}{(2\pi)^3}
\Bigl[3E_f 
+\frac{1}{\beta}\left(\ln{\cal F}_f+\ln{\cal F}_{\bar f}\right)\Bigl]
+U_{\rm M}(\sigma_f)+{\cal U}(\Phi,T), 
\label{PNJL-Omega}
\end{align}
%%%%%%%%%%%%%
where 
%%%%%%%%%%%%%
\begin{eqnarray}
{\cal F}_f=&1+3\Phi e^{-\beta E^-_f}+3\Phi^* e^{-2\beta E^-_f}
+e^{-3\beta E^-_f},\\
{\cal F}_{\bar f}=&1+3\Phi^* e^{-\beta E^+_f}+3\Phi e^{-2\beta E^+_f}
+e^{-3\beta E^+_f}. 
\label{factor_3_omega}
\end{eqnarray}
%%%%%%%%%%%%%
Note that ${\cal F}_2$ (${\cal F}_{\bar{2}}$) is the complex conjugate to 
${\cal F}_3$ (${\cal F}_{\bar{3}}$), indicating that $\Omega$ is real.

In the case of $\theta=2\pi/3$, particularly, 
$\Omega$ is invariant under the ${\mathbb Z}_3$ transformation,
%%%%%%%%%%%%%
\begin{eqnarray}
\Phi  \to e^{-i{2\pi k/{3}}} \Phi, \quad
\Phi^{*} \to e^{i{2\pi k/{3}}}\Phi^{*}.
\label{Z3}
\end{eqnarray}
%%%%%%%%%%%%%
Namely, $\Omega$ possesses the $\mathbb{Z}_3$ symmetry. 
When the exact color-confinement with $\Phi=0$ occurs, 
$\Omega$ is invariant for any interchange among $E_1^\pm$, $E_2^\pm$ and $E_3^\pm$. 
Namely, $\Omega$ has the flavor symmetry in the exact color-confinement 
phase.

%%%%%%%%%%%%%%%%%%%%%%%%%%%%%%
%\section{Numerical results}
%%%%%%%%%%%%%%%%%%%%%%%%%%%%%%

\begin{figure}[htbp]
\begin{center}
\includegraphics[width=0.3\textwidth,angle=-90]{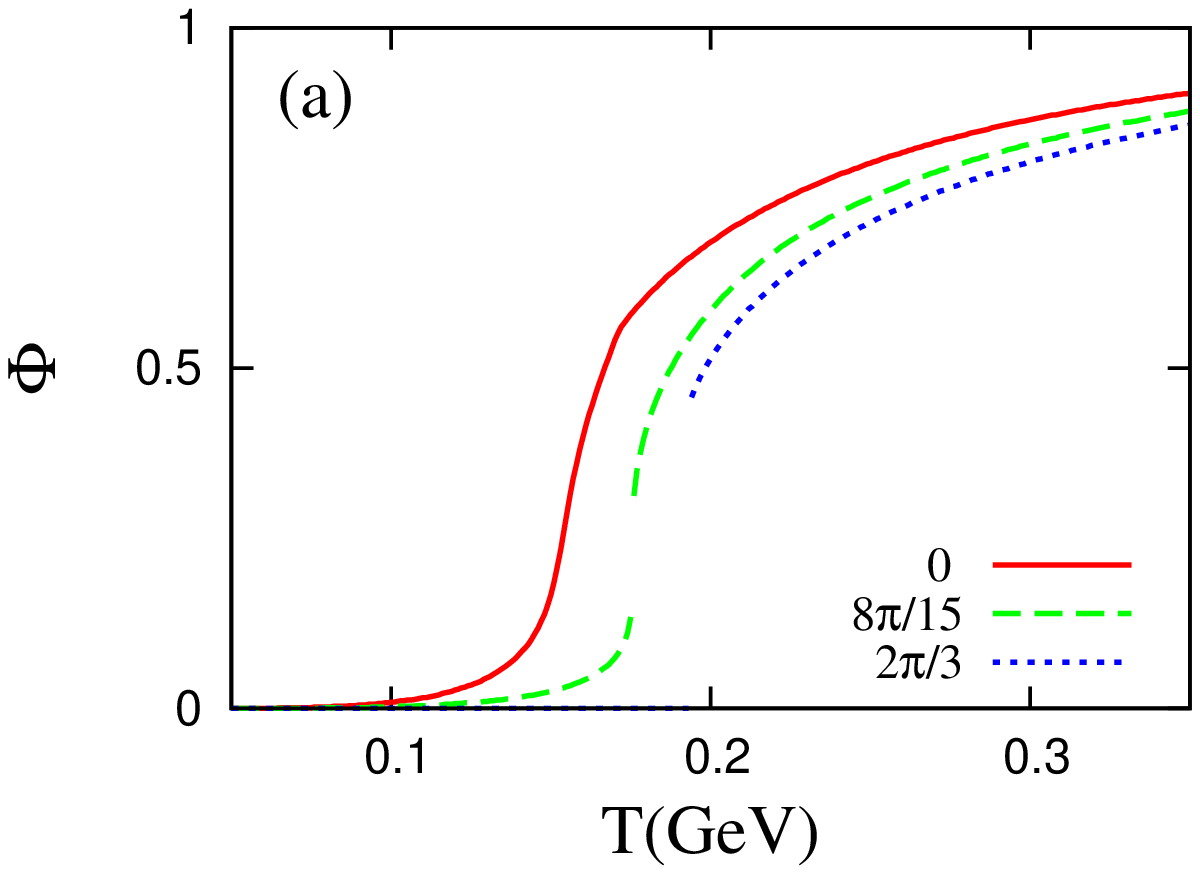}
\includegraphics[width=0.3\textwidth,angle=-90]{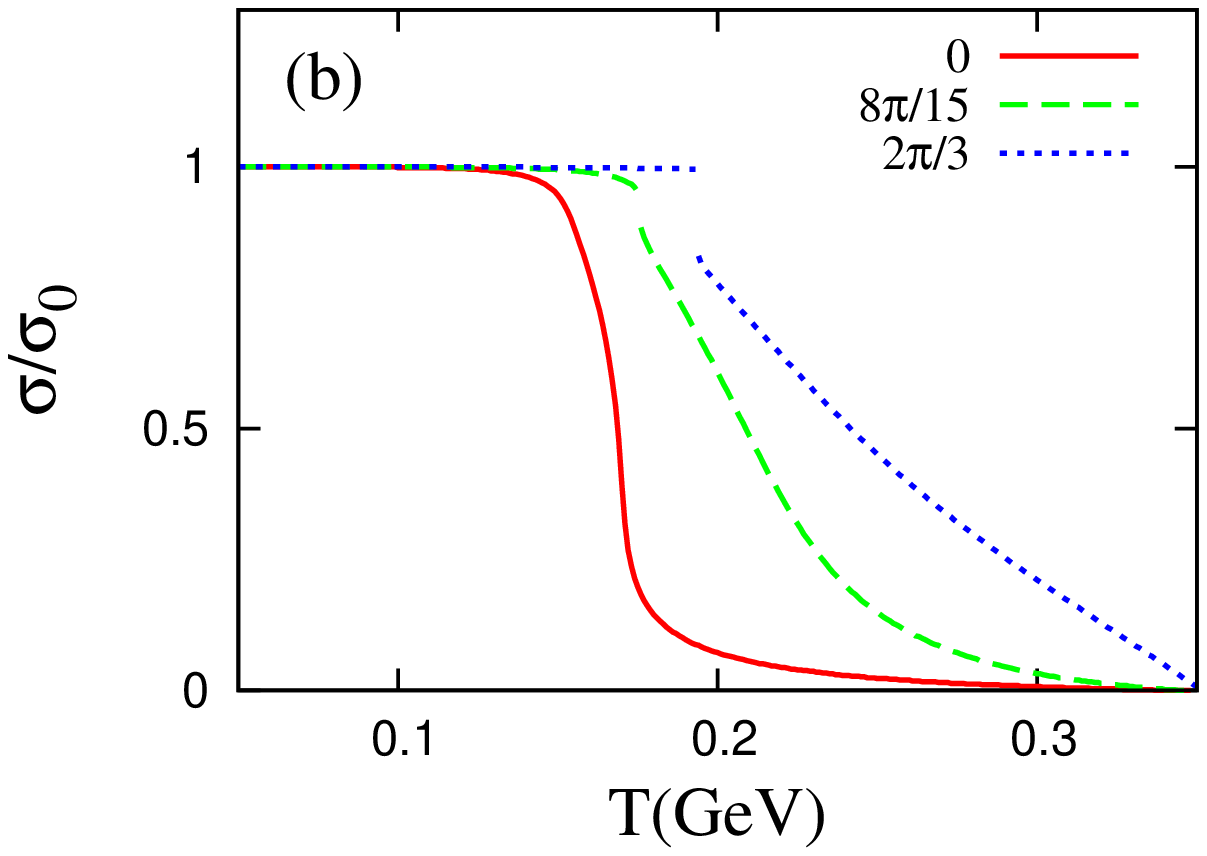}
\end{center}
\caption{(a) The Polyakov loop $\Phi$ and (b) the chiral condensate $\sigma_1$
in the $\theta$-$T$ plane at $\mu=0$MeV.
}
\label{Mu000}
\end{figure}

Figure \ref{Mu000} shows $T$ dependence of 
(a) the Polyakov loop $\Phi$ and (b) the chiral condensate $\sigma_1$
at $\mu=0$. 
The solid, dashed and dotted curves represent three cases of 
$\theta=0$, $8\pi/15$ and $2\pi/3$, respectively. 
For $\theta=0$ corresponding to the standard boundary condition, 
the chiral and deconfinement transitions are both crossover.
For $\theta=2\pi/3$ corresponding to the TBC, 
the first-order deconfinement transition occurs at $T=T_c=203$MeV and 
the exact color-confinement phase appears below $T_c$. 
The first-order transition of $\Phi$ at $T=T_c$ propagates to $\sigma_f$ 
as a discontinuity.
For $\theta\ne2\pi/3$, the deconfinement transition is no longer exact.
As $\theta$ decreases from $2\pi/3$ to zero, 
$T$ dependence of $\Phi$ becomes slower, 
and near $\theta = \pi/2$ the order of the deconfinement transition 
is changed from the first-order to crossover.

%%%%%%%%%%%%%%%
%%%%%%%%%%%%

\begin{figure}[htbp]
\begin{center}
\includegraphics[width=0.3\textwidth,angle=-90]{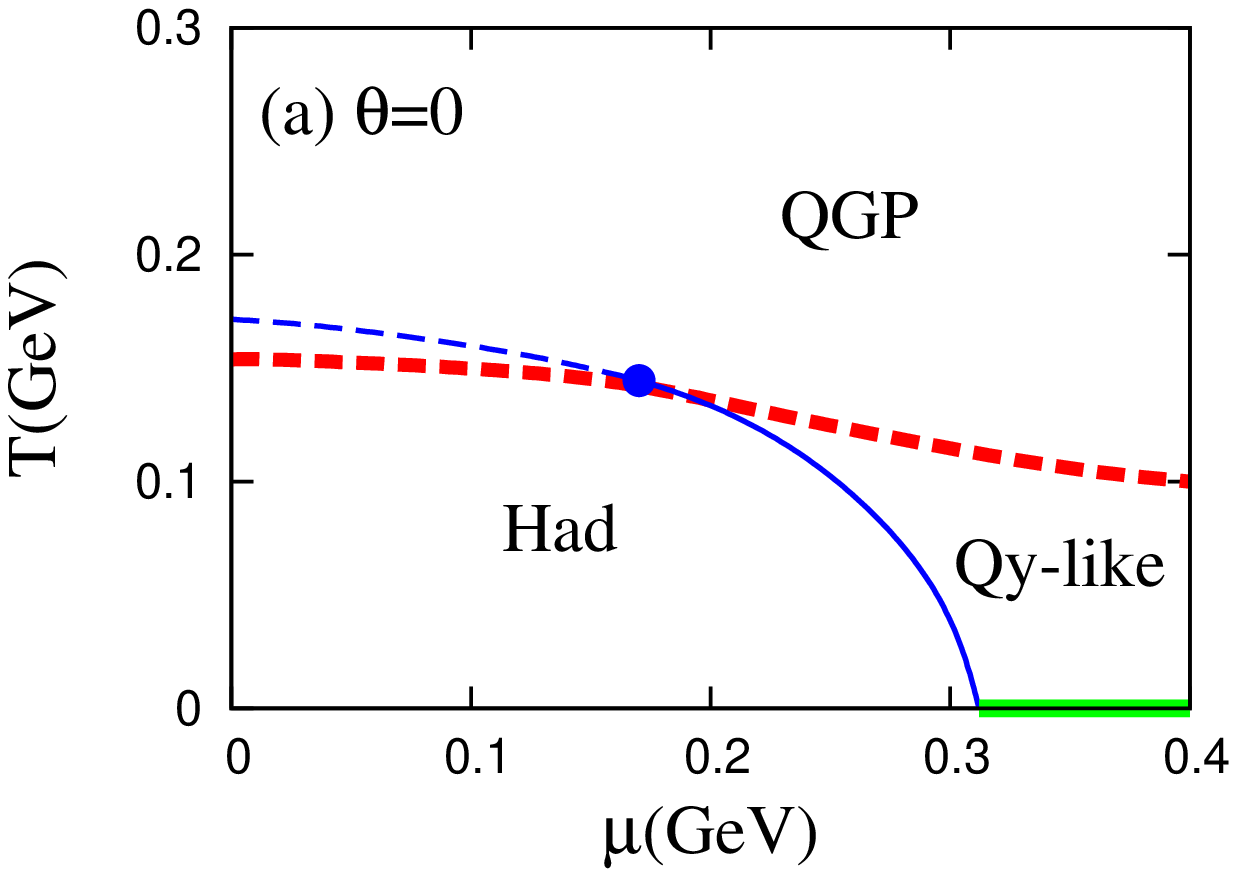}
\includegraphics[width=0.3\textwidth,angle=-90]{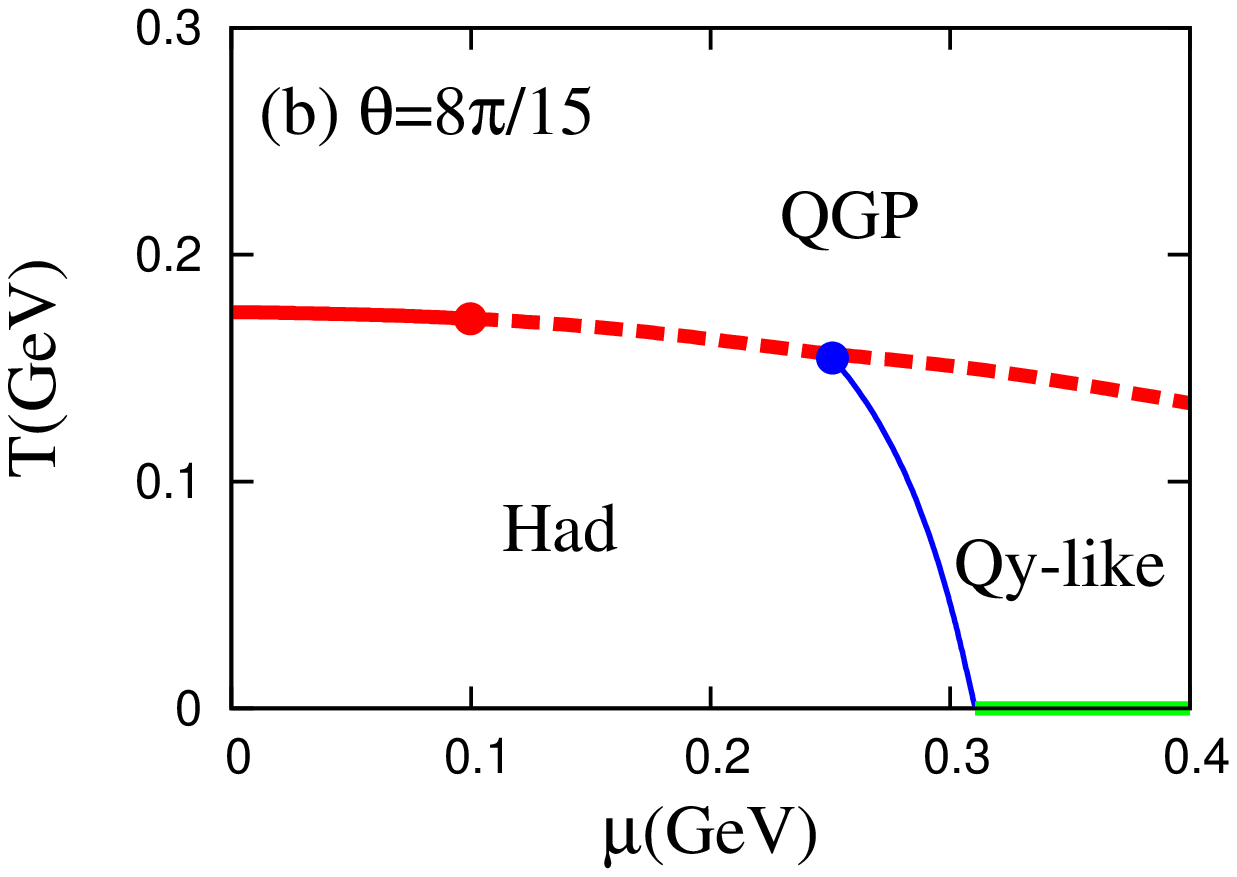}
\includegraphics[width=0.3\textwidth,angle=-90]{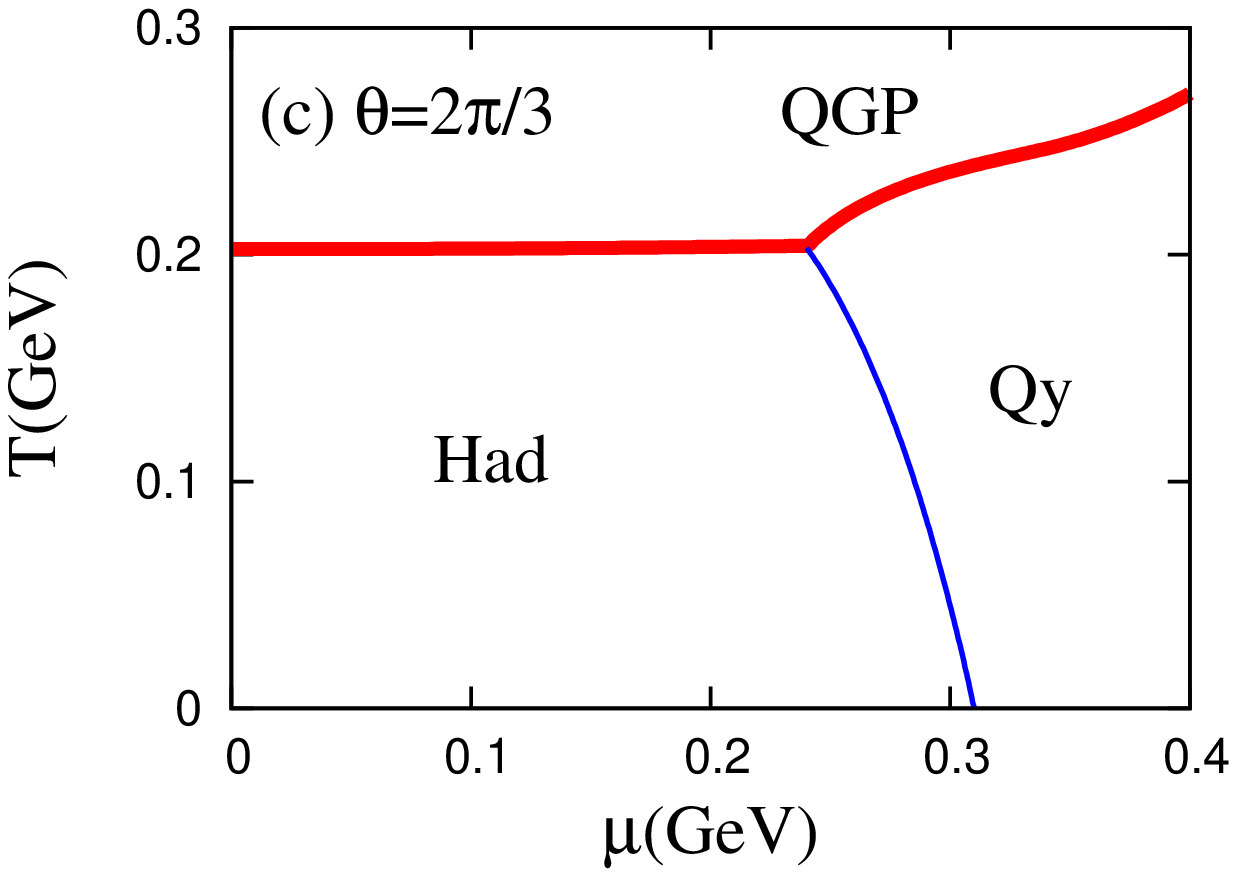}
\end{center}
\caption{
Phase diagram in the $T$-$\mu$ plane.
Panels (a)-(c) correspond to three cases 
of $\theta=0$, $8\pi/15$ and $2\pi/3$, respectively. 
The thick (thin) solid curve means the first-order deconfinement (chiral) 
phase transition line, 
while the thick (thin) dashed curve does the deconfinement (chiral) 
crossover line. The closed circles stand for the endpoints of the 
first-order deconfinement and chiral phase transition lines.
In panels (a) and (b), 
the thick-solid line at $T=0$ and $\mu \gtrsim M_f=323$~MeV 
represents the quarkyonic phase. 
}
\label{PD}
\end{figure}

Figure \ref{PD} shows the phase diagram in the $T$-$\mu$ plane. 
Panels (a)-(c) correspond to three cases 
of $\theta=0$, $8\pi/15$ and $2\pi/3$, respectively. 
The thick (thin) solid curve represents the first-order deconfinement (chiral) 
phase transition line, 
while the thick (thin) dashed curve corresponds to the deconfinement (chiral) 
crossover line defined by the peak of $d\Phi/dT$ ($d\sigma_f/dT$). 
For $\theta=0$, 
the chiral and deconfinement transitions are both crossover
at smaller $\mu$, but the former becomes the first-order at larger $\mu$.
For $\theta=2\pi/3$, the deconfinement transition is the 
first-order at any $\mu$, 
whereas the first-order chiral transition line appears only 
at $\mu \approx M_f=323$~MeV.
The region labeled by ``Qy" at $\mu \gtrsim M_f$ and small $T$ is 
the quarkyonic phase, since $\Phi=0$ and $n \neq 0$ there. 
The region labeled by ``Had" is the hadron phase, because 
the chiral symmetry is broken there and thereby the equation of state is 
dominated by the pion gas~\cite{Sakai_hadron}. 
The region labeled by ``QGP" corresponds to the quark gluon plasma (QGP) 
phase, although the flavor symmetry is broken there by the TBC. 
As $\theta$ decreases from $2\pi/3$ to zero, 
the first-order chiral transition line declines toward smaller $\mu$ and 
the critical endpoint moves to smaller $\mu$.
Once $\theta$ varies from $2\pi/3$, 
the quarkyonic phase defined by $\Phi=0$ and $n \neq 0$
shrinks on a line with $T=0$ and $\mu \gtrsim M_f$ and 
a region at small $T$ and $\mu \gtrsim M_f$ becomes 
a quarkyonic-like phase with small but finite $\Phi$ and $n \ne 0$; 
the latter region is labeled by ``Qy-like".

%%%%%%%%%%%%%%%
%%%%%%%%%%%%

For small $\theta$  far from $2\pi /3$, 
the deconfinement transition line declines 
as $\mu$ increases, but for $\theta =2\pi /3$ the line is almost horizontal 
at small $\mu$ and rises at intermediate $\mu$, as seen in Fig. \ref{PD}.   
The rising of the deconfinement transition line is a consequence of 
the $\mathbb{Z}_3$ symmetry, as shown below. 
The quark one-loop part of $\Omega$, 
which is defined by $\Omega_{\rm Q}$ in \eqref{PNJL-Omega_original-1}, 
can be expanded into a Maclaurin series
%%%%%%%%%%%%%
\begin{eqnarray}
\Omega_{\rm Q}=\Omega_{\rm Q}(\Phi =0,\Phi^*=0)+c_{10}\Phi +c_{01}\Phi^* +c_{20}\Phi^2+c_{11}\Phi\Phi^*+c_{02}{\Phi^*}^2+\cdots .
\label{GL}
\end{eqnarray}
The coefficients $c_{nm}$ are explicitly obtained as 
%%%%%%%%%%%%%
\begin{eqnarray}
c_{10}&=&-\frac{18}{\beta}\sum_f\int\frac{d^3{\bf p}}{(2\pi)^3}
\left(\frac{e^{-\beta E^{-}_f}}{1+e^{-3\beta E^{-}_f}}+
\frac{e^{-2\beta E^{+}_f}}{1+e^{-3\beta E^{+}_f}}\right)<0, 
\label{c10}\\
c_{01}&=&-\frac{18}{\beta}\sum_f\int\frac{d^3{\bf p}}{(2\pi)^3}
\left(\frac{e^{-2\beta E^{-}_f}}{1+e^{-3\beta E^{-}_f}}+
\frac{e^{-\beta E^{+}_f}}{1+e^{-3\beta E^{+}_f}}\right)<0, 
\label{c01}\\
c_{20}&=&\frac{9}{\beta}\sum_f\int\frac{d^3{\bf p}}{(2\pi)^3}
\left(\frac{e^{-2\beta E^{-}_f}}{(1+e^{-3\beta E^{-}_f})^2}+
\frac{e^{-4\beta E^{+}_f}}{(1+e^{-3\beta E^{+}_f})^2}\right)>0, 
\label{c20}\\
c_{11}&=&\frac{18}{\beta}\sum_f\int\frac{d^3{\bf p}}{(2\pi)^3}
\left(\frac{e^{-3\beta E^{-}_f}}{(1+e^{-3\beta E^{-}_f})^2}+
\frac{e^{-3\beta E^{+}_f}}{(1+e^{-3\beta E^{+}_f})^2}\right)>0, 
\label{c11}\\
c_{02}&=&\frac{9}{\beta}\sum_f\int\frac{d^3{\bf p}}{(2\pi)^3}
\left(\frac{e^{-4\beta E^{-}_f}}{(1+e^{-3\beta E^{-}_f})^2}+
\frac{e^{-2\beta E^{+}_f}}{(1+e^{-3\beta E^{+}_f})^2}\right)>0. 
\label{c02}
\end{eqnarray}
%%%%%%%%%%%%%
The $c_{nm}$ are positive for even $n+m$ but negative for odd $n+m$. 
The absolute values of the $c_{nm}$ increase as $\mu$ increases, 
unless $\mu$ is quite large. 
For simplicity, we fix $M_f$ to a constant to focus 
our attention on $\Phi$ dependence of $\Omega$. 
In this assumption, $U_{\rm M}$ and the zeroth-order term 
$\Omega_{\rm Q}(\Phi =0,\Phi^*=0)$ in the  Maclaurin series 
are just constants and thereby become irrelevant to the present discussion. 
So we neglect these terms. 
We also assume that 
$\Phi=\Phi^*$. This is true for $\mu=0$ and well satisfied for 
small and intermediate $\mu$ of our interest. 
In the pure gauge limit where $\Omega_{\rm Q}=0$, the thermodynamic potential 
$\Omega$ agrees with 
the Polyakov potential ${\cal U}(\Phi)$ and hence has no $\mu$ dependence. 
The potential has a global minimum 
at $\Phi=0$ and a local one at $\Phi=\Phi_{m}>0$ for small $T$: namely, 
${\cal U}(\Phi=0) < {\cal U}(\Phi=\Phi_{m})$. 
For the case of $\theta=2\pi/3$, the system has the $\mathbb{Z}_3$ symmetry. 
Up to the second order of the  Maclaurin series, only the $c_{11}\Phi\Phi^*$ 
term appears because of the symmetry. When the term is added to ${\cal U}$, 
the resultant potential keeps the same value as ${\cal U}(\Phi)$ at 
$\Phi=0$, but increases from ${\cal U}(\Phi)$ at $\Phi >0$.  
This property makes the deconfinement transition more difficult. 
The coefficient $c_{11}$ as a function of $\mu$ little increases 
for $\mu \ll M_f$, but the increase becomes sizable for $\mu > M_f$. 
Therefore the rising of the deconfinement transition line 
with respect to increasing $\mu$ is tiny at small $\mu$ but becomes 
sizable at intermediate $\mu$. 
The potential $c_{11}\Phi\Phi^*+{\cal U}$ keeps 
a positive curvature at $\Phi=0$ because of $c_{11}>0$, 
so that the deconfinement transition is first-order for 
any positive value of $c_{11}$. 
For the case of $\theta \neq 2\pi/3$, meanwhile, 
the first-order term $c_{10}\Phi +c_{01}\Phi^*$ is not prohibited by 
the $\mathbb{Z}_3$ symmetry and thereby dominates $\Omega_{\rm Q}$ 
particularly for small $\theta$ far from $2\pi/3$. 
Since $c_{10}$ and $c_{01}$ are negative, 
the situation for small $\theta$ becomes opposite to 
that for $\theta = 2\pi/3$. 
Eventually, the deconfinement transition line slopes down as 
$\mu$ increases for the case of small $\theta$.

\begin{figure}[htbp]
\begin{center}
\includegraphics[width=0.4\textwidth]{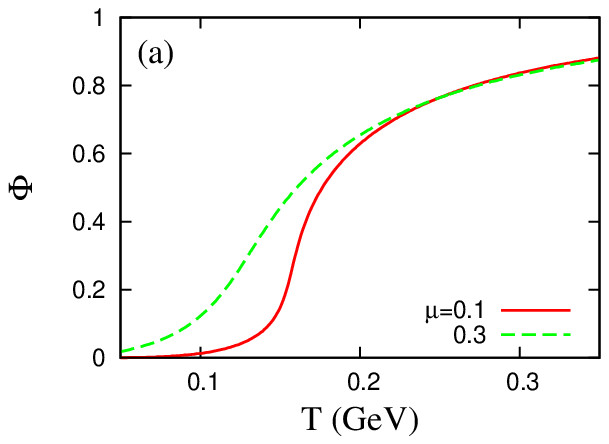}
\includegraphics[width=0.4\textwidth]{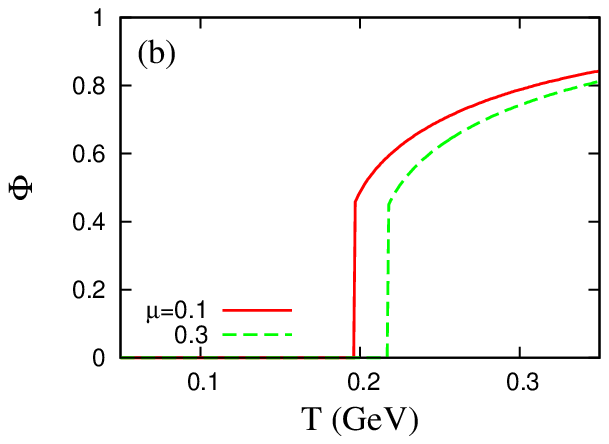}
\end{center}
\caption{$T$ dependence of $\Phi$ for $\mu=0.1,~0.3$ GeV 
in the lowest order approximation. 
Panel (a) corresponds to the case of $\theta =0$ and panel (b) does to the 
case of $\theta =2\pi /3$. 
}
\label{F}
\end{figure}

Figure \ref{F} shows $T$ dependence of $\Phi$ for 
$\mu=0.1, 0.3$ GeV. We consider the case of $\theta=0$ in panel (a) and 
that of $\theta=2\pi/3$ in panel (b) by assuming 
$\Omega=c_{10}\Phi +c_{01}\Phi^* + {\cal U}$ in panel (a) and 
$\Omega=c_{11}\Phi \Phi^* + {\cal U}$ in panel (b). 
Note that $M_f$ is fixed to 323 MeV and 
$T$ dependence of $\Phi$ is determined from the $\Omega$ 
with the minimum condition. 
As $\mu$ increases, the transition temperature decreases 
for $\theta=0$, but increases for $\theta=2\pi/3$. 
The transition is first-order for the case of $\theta=2\pi/3$. 
These results are consistent with the qualitative discussion mentioned above.

%%%%%%%%%%%%%%%
%%%%%%%%%%%%

%%%%%%%%%%%%%%%%%%%%%%%%%%%%%%%%%%%%%%%%%%%%%%%%%%
%%%%%%%%  Summary 
%%%%%%%%%%%%%%%%%%%%%%%%%%%%%%%%%%%%%%%%%%%%%%%%%%

%\section{Summary}
% \label{Summary}

In summary, we have investigated 
the interplay between the $\mathbb{Z}_{N_c}$ symmetry and the emergence of 
the quarkyonic phase, adding the complex chemical potentials 
$\mu_f=\mu+i T \theta_f$ with $(\theta_f)=(0,\theta,-\theta)$ to 
the PNJL model. 
When $\theta=0$, 
the PNJL model with the $\mu_f$ is reduced to the PNJL model with real $\mu$. 
This situation corresponds to QCD at real $\mu$. 
When $\theta=2\pi/3$, meanwhile, 
the PNJL model with the $\mu_f$ is reduced to the TBC model 
with the $\mathbb{Z}_{N_c}$ symmetry. 
This situation corresponds to the QCD-like theory with 
the $\mathbb{Z}_{N_c}$ symmetry at real $\mu$. 
When $\theta=2\pi/3$, the quarkyonic phase defined by $\Phi=0$ and $n > 0$ 
really exists at small $T$ and large $\mu$. 
Once $\theta$ varies from $2\pi/3$ to zero, 
the $\mathbb{Z}_{N_c}$ symmetry is broken. 
As a consequence of this property, the quarkyonic phase exists only 
on a line of $T=0$ and $\mu \gtrsim M_f$, and the region 
at small $T$ and large $\mu$ is dominated by 
the quarkyonic-like phase characterized by 
small but finite $\Phi$ and $n>0$. 
The $\mathbb{Z}_{N_c}$ symmetry thus plays an essential role on 
the emergence and the location of the quarkyonic phase 
in the $\mu$-$T$ plane, and the quarkyonic-like phase at $\theta=0$ 
is a remnant of the quarkyonic phase at $\theta=2\pi/3$.  
Since the $\mathbb{Z}_{N_c}$ symmetry is explicitly broken at 
$\theta=0$, it is then natural to expand the concept of 
the quarkyonic phase and redefine it by a phase with small $\Phi$ and 
finite $n$. For this reason, the quarkyonic-like phase is 
often called the quarkyonic phase. 
The gross structure of the phase diagram thus has no qualitative 
difference between $\theta=2\pi/3$ and zero, 
if the concept of the quarkyonic phase is properly expanded. 
In this sense,  
the $\mathbb{Z}_{N_c}$ symmetry is a good approximate concept 
for the case of $\theta=0$, even if the current quark mass is small.

\noindent
\begin{acknowledgments}
The authors thank A. Nakamura, T. Saito, K. Nagata and K. Kashiwa for useful discussions. 
H.K. also thanks M. Imachi, H. Yoneyama, H. Aoki and M. Tachibana for useful discussions. 
Y.S. is supported by RIKEN Special Postdoctoral Researchers Program.
T.S. are supported by JSPS. 
\end{acknowledgments}

%%%%%%%%%%%%%%%%%%%%%%%%%%%%%%%%%%%%%%%%%%%%%%%%%%%%%%%%%%%%%%%%%%%%%%%%%%%%%%%%%%%%% References 
%%%%%%%%%%%%%%%%%%%%%%%%%%%%%%%%%%%%%%%%%%%%%%%%%%%%%%%%%%%%%%%%%%%%%%%%%%%%%%%%


\begin{thebibliography}{19}
\expandafter\ifx\csname natexlab\endcsname\relax\def\natexlab#1{#1}\fi
\expandafter\ifx\csname bibnamefont\endcsname\relax
  \def\bibnamefont#1{#1}\fi
\expandafter\ifx\csname bibfnamefont\endcsname\relax
  \def\bibfnamefont#1{#1}\fi
\expandafter\ifx\csname citenamefont\endcsname\relax
  \def\citenamefont#1{#1}\fi
\expandafter\ifx\csname url\endcsname\relax
  \def\url#1{\texttt{#1}}\fi
\expandafter\ifx\csname urlprefix\endcsname\relax\def\urlprefix{URL }\fi
\providecommand{\bibinfo}[2]{#2}
\providecommand{\eprint}[2][]{\url{#2}}
%
%%%%%%%%%%%%%%%%%%%%%%%%%%%%%%%%%%%%%%%%%%%%%%%%%%
%
%%%%%%%%%%%%%%%%%%%%%%%%%%%%%%%%%%%%%%%%%%%%%%%%%%%%%%%%%%
% [1] 
%%%%%%%%%%%%%%%%%%%%%%%%%%%%%%%%%%%%%%%%%%%%%%%%%%%%%%%%%%
\bibitem[{\citenamefont{Kouno et al}(2012)}]{Kouno_TBC}
\bibinfo{author}{\bibfnamefont{H.}~\bibnamefont{Kouno}}, 
\bibinfo{author}{\bibfnamefont{Y.}~\bibnamefont{Sakai}}, 
\bibinfo{author}{\bibfnamefont{T.}~\bibnamefont{Makiyama}},
\bibinfo{author}{\bibfnamefont{K.}~\bibnamefont{Tokunaga}},
\bibinfo{author}{\bibfnamefont{T.}~\bibnamefont{Sasaki}},   
\bibnamefont{and}
\bibinfo{author}{\bibfnamefont{M.}~\bibnamefont{Yahiro}}, 
\bibinfo{journal}{J. Phys. G: Nucl. Part. Phys. } \textbf{\bibinfo{volume}{39}},
\bibinfo{pages}{085010} (\bibinfo{year}{2012}). 
%\bibinfo{howpublished}{arXiv:1202.5584 [hep-ph]}(\bibinfo{year}{2012}). 
%%%%%%%%%%%%%%%%%%%%%%%%%%%%
%
%%%%%%%%%%%%%%%%%%%%%%%%%%%%%%%%%%%%%%%%%%%%%%%%%%
% [2] 
%%%%%%%%%%%%%%%%%%%%%%%%%%%%%%%%%%%%%%%%%%%%%%%%%%
\bibitem[{\citenamefont{Roberge and Weiss}(1986)}]{RW}
\bibinfo{author}{\bibfnamefont{A.}~\bibnamefont{Roberge}} \bibnamefont{and}
\bibinfo{author}{\bibfnamefont{N.}~\bibnamefont{Weiss}},  
\bibinfo{journal}{Nucl. Phys. } \textbf{\bibinfo{volume}{B275}},
\bibinfo{pages}{734} (\bibinfo{year}{1986}). 
%%%%%%%%%%%%%%%%%%%%%%%%%%%%%%%%%%%%%%%%%%%%%%%%%%
%
%%%%%%%%%%%%%%%%%%%%%%%%%%%%%%%%%%%%%%%%%%%%%%%%%%
% [3] 
%%%%%%%%%%%%%%%%%%%%%%%%%%%%%%%%%%%%%%%%%%%%%%%%%%
\bibitem[{\citenamefont{Sakai et al}(2008)}]{Sakai}
\bibinfo{author}{\bibfnamefont{Y.}~\bibnamefont{Sakai}},
\bibinfo{author}{\bibfnamefont{K.}~\bibnamefont{Kashiwa}}, 
\bibinfo{author}{\bibfnamefont{H.}~\bibnamefont{Kouno}}, 
\bibnamefont{and}
\bibinfo{author}{\bibfnamefont{M.}~\bibnamefont{Yahiro}},
\bibinfo{journal}{Phys.\ Rev.\  D} \textbf{\bibinfo{volume}{77}},
\bibinfo{pages}{051901(R)} (\bibinfo{year}{2008}). 
%%%%%%%%%%%%%%%%%%%%%%%%%%%%%%%%%%%%%%%%%%%%%%%%%%
%
%%%%%%%%%%%%%%%%%%%%%%%%%%%%%%%%%%%%%%%%%%%%%%%%%%
% [4] 
%%%%%%%%%%%%%%%%%%%%%%%%%%%%%%%%%%%%%%%%%%%%%%%%%%
\bibitem[{\citenamefont{Meisinger et al.}(1996)}]{Meisinger}
\bibinfo{author}{\bibfnamefont{P.}~\bibnamefont{N.}}~\bibnamefont{Meisinger},
\bibnamefont{and}
\bibinfo{author}{\bibfnamefont{M.}~\bibnamefont{C.}}~\bibnamefont{Ogilvie},  
  \bibinfo{journal}{Phys. Lett.\ B} \textbf{\bibinfo{volume}{379}},
  \bibinfo{pages}{163} (\bibinfo{year}{1996}). 
%%%%%%%%%%%%%%%%%%%%%%%%%%%%%%%%%%%%%%%%%%%%%%%%%%
%
%%%%%%%%%%%%%%%%%%%%%%%%%%%%%%%%%%%%%%%%%%%%%%%%%%%%%%%%%
% [5] 
%%%%%%%%%%%%%%%%%%%%%%%%%%%%%%%%%%%%%%%%%%%%%%%%%%%%%%%%%
\bibitem[{\citenamefont{Dumitru}(2002)}]{Dumitru}
\bibinfo{author}{\bibfnamefont{A.}~\bibnamefont{Dumitru}},
\bibnamefont{and}
\bibinfo{author}{\bibfnamefont{R.}~\bibfnamefont{D.}~\bibnamefont{Pisarski}},  
\bibinfo{journal}{Phys.\ Rev.\  D} \textbf{\bibinfo{volume}{66}},
\bibinfo{pages}{096003} (\bibinfo{year}{2002}); 
\bibinfo{author}{\bibfnamefont{A.}~\bibnamefont{Dumitru}},
\bibinfo{author}{\bibfnamefont{Y.}~\bibnamefont{Hatta}},
\bibinfo{author}{\bibfnamefont{J.}~\bibnamefont{Lenaghan}},
\bibinfo{author}{\bibfnamefont{K.}~\bibnamefont{Orginos}},
\bibnamefont{and}
\bibinfo{author}{\bibfnamefont{R.}~\bibfnamefont{D.}~\bibnamefont{Pisarski}},  
\bibinfo{journal}{Phys.\ Rev.\  D} \textbf{\bibinfo{volume}{70}},
\bibinfo{pages}{034511} (\bibinfo{year}{2004}); 
\bibinfo{author}{\bibfnamefont{A.}~\bibnamefont{Dumitru}},
\bibinfo{author}{\bibfnamefont{R.}~\bibfnamefont{D.}~\bibnamefont{Pisarski}},  
\bibnamefont{and}
\bibinfo{author}{\bibfnamefont{D.}~\bibnamefont{Zschiesche}},  
\bibinfo{journal}{Phys.\ Rev.\  D} \textbf{\bibinfo{volume}{72}},
\bibinfo{pages}{065008} (\bibinfo{year}{2005}).
%%%%%%%%%%%%%%%%%%%%%%%%%%%%%%%%%%%%%%%%%%%%%%%
%
%%%%%%%%%%%%%%%%%%%%%%%%%%%%%%%%%%%%%%%%%%%%%%%%%%
% [6] 
%%%%%%%%%%%%%%%%%%%%%%%%%%%%%%%%%%%%%%%%%%%%%%%%%%
\bibitem[{\citenamefont{Fukushima}(2004)}]{Fukushima}
\bibinfo{author}{\bibfnamefont{K.}~\bibnamefont{Fukushima}}, 
  \bibinfo{journal}{Phys. Lett.\ B} \textbf{\bibinfo{volume}{591}},
  \bibinfo{pages}{277} (\bibinfo{year}{2004}).. 
%%%%%%%%%%%%%%%%%%%%%%%%%%%%%%%%%%%%%%%%%%%%%%%%%%
%
%%%%%%%%%%%%%%%%%%%%%%%%%%%%%%%%%%%%%%%%%%%%%%%%%%%%%
% [7]
%%%%%%%%%%%%%%%%%%%%%%%%%%%%%%%%%%%%%%%%%%%%%%%%%%%%%%
\bibitem[{\citenamefont{Ratti et al.}(2006)}]{Ratti}
\bibinfo{author}{\bibfnamefont{C.}~\bibnamefont{Ratti}},
\bibinfo{author}{\bibfnamefont{M.}~\bibfnamefont{A.}~\bibnamefont{Thaler}},
\bibnamefont{and}
\bibinfo{author}{\bibfnamefont{W.}~\bibnamefont{Weise}},  
  \bibinfo{journal}{Phys. Rev.\ D} \textbf{\bibinfo{volume}{73}},
  \bibinfo{pages}{014019} (\bibinfo{year}{2006}); 
\bibinfo{author}{\bibfnamefont{C.}~\bibnamefont{Ratti}},
\bibinfo{author}{\bibfnamefont{S.}~\bibnamefont{R\"{o}{\ss}ner}},
\bibinfo{author}{\bibfnamefont{M.}~\bibfnamefont{A.}~\bibnamefont{Thaler}},
\bibnamefont{and}
\bibinfo{author}{\bibfnamefont{W.}~\bibnamefont{Weise}},  
  \bibinfo{journal}{Eur. Phys. J.\ C} \textbf{\bibinfo{volume}{49}},
  \bibinfo{pages}{213} (\bibinfo{year}{2007}). 
%%%%%%%%%%%%%%%%%%%%%%%%%%%%%%%%%%%%%%%%%%%%%%%
%
%%%%%%%%%%%%%%%%%%%%%%%%%%%%%%%%%%%%%%%%%%%%%%%%%%
% [8] 
%%%%%%%%%%%%%%%%%%%%%%%%%%%%%%%%%%%%%%%%%%%%%%%%%%
\bibitem[{\citenamefont{Rossner et al.}(2007)}]{Rossner}
\bibinfo{author}{\bibfnamefont{S.}~\bibnamefont{R\"{o}{\ss}ner}},
\bibinfo{author}{\bibfnamefont{C.}~\bibnamefont{Ratti}},
\bibnamefont{and}
\bibinfo{author}{\bibfnamefont{W.}~\bibnamefont{Weise}},  
  \bibinfo{journal}{Phys. Rev.\ D} \textbf{\bibinfo{volume}{75}},
  \bibinfo{pages}{034007} (\bibinfo{year}{2007}). 
%%%%%%%%%%%%%%%%%%%%%%%%%%%%%%%%%%%%%%%%%%%%%%%%%%
%
%%%%%%%%%%%%%%%%%%%%%%%%%%%%%%%%%%%%%%%%%%%%%%%%%%
% [9] 
%%%%%%%%%%%%%%%%%%%%%%%%%%%%%%%%%%%%%%%%%%%%%%%%%%
\bibitem[{\citenamefont{Schaefer}(2007)}]{Schaefer}
\bibinfo{author}{\bibfnamefont{B.}~\bibfnamefont{-J.}~\bibnamefont{Schaefer}},
\bibinfo{author}{\bibfnamefont{J.}~\bibfnamefont{M.}~\bibnamefont{Pawlowski}},
\bibnamefont{and}
\bibinfo{author}{\bibfnamefont{J.}~\bibnamefont{Wambach}},  
  \bibinfo{journal}{Phys.\ Rev.\  D} \textbf{\bibinfo{volume}{76}},
  \bibinfo{pages}{074023} (\bibinfo{year}{2007}).
%%%%%%%%%%%%%%%%%%%%%%%%%%%%%%%%%%%%%%%%%%%%%%%%%%
%
%%%%%%%%%%%%%%%%%%%%%%%%%%%%%%%%%%%%%%%%%%%%%%%%%%
% [10] 
%%%%%%%%%%%%%%%%%%%%%%%%%%%%%%%%%%%%%%%%%%%%%%%%%%%
\bibitem[{\citenamefont{Abuki et al}(2008)}]{Abuki}
\bibinfo{author}{\bibfnamefont{H.}~\bibnamefont{Abuki}},
\bibinfo{author}{\bibfnamefont{R.}~\bibnamefont{Anglani}},
\bibinfo{author}{\bibfnamefont{R.}~\bibnamefont{Gatto}},
\bibinfo{author}{\bibfnamefont{G.}~\bibnamefont{Nardulli}},
\bibnamefont{and}
\bibinfo{author}{\bibfnamefont{M.}~\bibnamefont{Ruggieri}},
\bibinfo{journal}{Phys.\ Rev.\  D} \textbf{\bibinfo{volume}{78}},
\bibinfo{pages}{034034} (\bibinfo{year}{2008}). 
%%%%%%%%%%%%%%%%%%%%%%%%%%%%%%%%%%%%%%%%%%%%%%%%%
%
%%%%%%%%%%%%%%%%%%%%%%%%%%%%%%%%%%%%%%%%%%%%%%%%%%%%
% [11] 
%%%%%%%%%%%%%%%%%%%%%%%%%%%%%%%%%%%%%%%%%%%%%%%%%%%%%
\bibitem[{\citenamefont{Fukushima}(2004)}]{Fukushima2}
\bibinfo{author}{\bibfnamefont{K.}~\bibnamefont{Fukushima}}, 
  \bibinfo{journal}{Phys.\ Rev.\  D} \textbf{\bibinfo{volume}{77}},
  \bibinfo{pages}{114028} (\bibinfo{year}{2008}).
%%%%%%%%%%%%%%%%%%%%%%%%%%%%%%%%%%%%%%%%%%%%%%%
%
%%%%%%%%%%%%%%%%%%%%%%%%%%%%%%%%%%%%%%%%%%%%%%%%%%
% [12] 
%%%%%%%%%%%%%%%%%%%%%%%%%%%%%%%%%%%%%%%%%%%%%%%%%%
\bibitem[{\citenamefont{Kashiwa et al}(2008)}]{Kashiwa1}
\bibinfo{author}{\bibfnamefont{K.}~\bibnamefont{Kashiwa}}, 
\bibinfo{author}{\bibfnamefont{H.}~\bibnamefont{Kouno}}, 
\bibinfo{author}{\bibfnamefont{M.}~\bibnamefont{Matsuzaki}}, 
\bibnamefont{and}
\bibinfo{author}{\bibfnamefont{M.}~\bibnamefont{Yahiro}},
  \bibinfo{journal}{Phys.\ Lett.\ B} \textbf{\bibinfo{volume}{662}},
  \bibinfo{pages}{26} (\bibinfo{year}{2008}).
%%%%%%%%%%%%%%%%%%%%%%%%%%%%%%%%%%%%%%%%%%%%%%%%%%
%
%%%%%%%%%%%%%%%%%%%%%%%%%%%%%%%%%%%%%%%%%%%%%%%%%%%%%%%%
% [13] 
%%%%%%%%%%%%%%%%%%%%%%%%%%%%%%%%%%%%%%%%%%%%%%%%%%%%%%%%
\bibitem{McLerran_largeNc}
\bibinfo{author}{\bibfnamefont{L.}~\bibnamefont{McLerran}}
\bibinfo{author}{\bibfnamefont{K.}~\bibnamefont{Redlich}}
\bibnamefont{and}
\bibinfo{author}{\bibfnamefont{C.}~\bibnamefont{Sasaki}}, 
  \bibinfo{journal}{Nucl. Phys.\  A} \textbf{\bibinfo{volume}{824}},
  \bibinfo{pages}{86} (\bibinfo{year}{2009}).
%%%%%%%%%%%%%%%%%%%%%%%%%%%%%%%%%%%%%%%%%%%%%
%
%%%%%%%%%%%%%%%%%%%%%%%%%%%%%%%%%%%%%%%%%%%%%%%%%%%%%
% [14]
%%%%%%%%%%%%%%%%%%%%%%%%%%%%%%%%%%%%%%%%%%%%%%%%%%%%%%
\bibitem[{\citenamefont{Hell et al.}(2010)}]{Hell}
\bibinfo{author}{\bibfnamefont{T.}~\bibnamefont{Hell}},
\bibinfo{author}{\bibfnamefont{S.}~\bibnamefont{R\"{o}{\ss}ner}},
\bibinfo{author}{\bibfnamefont{M.}~\bibnamefont{Cristoforetti}},
\bibnamefont{and}
\bibinfo{author}{\bibfnamefont{W.}~\bibnamefont{Weise}},
\bibinfo{journal}{Phys. Rev.\ D} \textbf{\bibinfo{volume}{81}},
\bibinfo{pages}{074034} (\bibinfo{year}{2010}); 
\bibinfo{author}{\bibfnamefont{T.}~\bibnamefont{Hell}},
\bibinfo{author}{\bibfnamefont{K.}~\bibnamefont{Kashiwa}},
\bibnamefont{and}
\bibinfo{author}{\bibfnamefont{W.}~\bibnamefont{Weise}},
\bibinfo{journal}{Phys. Rev.\ D} \textbf{\bibinfo{volume}{83}},
\bibinfo{pages}{114008} (\bibinfo{year}{2011}).
%%%%%%%%%%%%%%%%%%%%%%%%%%%%%%%%%%%%%%%%%%%%%%%
%
%%%%%%%%%%%%%%%%%%%%%%%%%%%%%%%%%%%%%%%%%%%%%%%%%%
% [15] 
%%%%%%%%%%%%%%%%%%%%%%%%%%%%%%%%%%%%%%%%%%%%%%%%%%
\bibitem[{\citenamefont{Sakai et al.}(2010)}]{Sakai_imiso}
\bibinfo{author}{\bibfnamefont{Y.}~\bibnamefont{Sakai}},
\bibinfo{author}{\bibfnamefont{H.}~\bibnamefont{Kouno}},
\bibnamefont{and}
\bibinfo{author}{\bibfnamefont{M.}~\bibnamefont{Yahiro}},
\bibinfo{journal}{J. Phys. \  G: Nucl. Part. Phys.} 
\textbf{\bibinfo{volume}{37}},
\bibinfo{pages}{105007} (\bibinfo{year}{2010}). 
%%%%%%%%%%%%%%%%%%%%%%%%%%%%%%%%%%%%%%%%%%%%%%%%%%
%
%%%%%%%%%%%%%%%%%%%%%%%%%%%%%%%%%%%%%%%%%%%%%%%%%%
% [16]
%%%%%%%%%%%%%%%%%%%%%%%%%%%%%%%%%%%%%%%%%%%%%%%%%%
\bibitem[{\citenamefont{Matsumoto}(2010)}]{Matsumoto}
\bibinfo{author}{\bibfnamefont{T.}~\bibnamefont{Matsumoto}}, 
\bibinfo{author}{\bibfnamefont{K.}~\bibnamefont{Kashiwa}}, 
\bibinfo{author}{\bibfnamefont{H.}~\bibnamefont{Kouno}}, 
\bibinfo{author}{\bibfnamefont{K.}~\bibnamefont{Oda}}, 
\bibnamefont{and} 
\bibinfo{author}{\bibfnamefont{M.}~\bibnamefont{Yahiro}}, 
\bibinfo{journal}{Phys. Lett. \  B} 
\textbf{\bibinfo{volume}{694}},
\bibinfo{pages}{367} (\bibinfo{year}{2011}).
%\bibinfo{howpublished}{arXiv:hep-ph/1004.0592 [hep-ph]} (\bibinfo{year}{2010}).
%%%%%%%%%%%%%%%%%%%%%%%%%%%%%%%%%%%%%%%%%%%%%%%%%%
%
%%%%%%%%%%%%%%%%%%%%%%%%%%%%%%%%%%%%%%%%%%%%%%%%%%
% [17] 
%%%%%%%%%%%%%%%%%%%%%%%%%%%%%%%%%%%%%%%%%%%%%%%%%%
\bibitem[{\citenamefont{Sakai}(2010)}]{Sakai5}
\bibinfo{author}{\bibfnamefont{Y.}~\bibnamefont{Sakai}},
\bibinfo{author}{\bibfnamefont{T.}~\bibnamefont{Sasaki}}, 
\bibinfo{author}{\bibfnamefont{H.}~\bibnamefont{Kouno}},
\bibnamefont{and}
\bibinfo{author}{\bibfnamefont{M.}~\bibnamefont{Yahiro}}, 
\bibinfo{journal}{Phys. Rev. \ D} 
\textbf{\bibinfo{volume}{82}},
\bibinfo{pages}{076003} (\bibinfo{year}{2010}). 
%%%%%%%%%%%%%%%%%%%%%%%%%%%%%%%%%%%%%%%%%%%%%%%%%%
%
%%%%%%%%%%%%%%%%%%%%%%%%%%%%%%%%%%%%%%%%%%%%%%%%%%
% [18] 
%%%%%%%%%%%%%%%%%%%%%%%%%%%%%%%%%%%%%%%%%%%%%%%%%%
\bibitem[{\citenamefont{Gatto}(2011)}]{Gatto}
\bibinfo{author}{\bibfnamefont{R.}~\bibnamefont{Gatto}},
\bibnamefont{and}
\bibinfo{author}{\bibfnamefont{M.}~\bibnamefont{Ruggieri}}, 
\bibinfo{journal}{Phys. Rev. \ D} 
\textbf{\bibinfo{volume}{83}},
\bibinfo{pages}{034016} (\bibinfo{year}{2011}).
%%%%%%%%%%%%%%%%%%%%%%%%%%%%%%%%%%%%%%%%%%%%%%%%%%
%
%%%%%%%%%%%%%%%%%%%%%%%%%%%%%%%%%%%%%%%%%%%%%%%%%%
% [19] 
%%%%%%%%%%%%%%%%%%%%%%%%%%%%%%%%%%%%%%%%%%%%%%%%%%
\bibitem[{\citenamefont{Sasaki et al.}(2011)}]{Sasaki-T_Nf3}
\bibinfo{author}{\bibfnamefont{T.}~\bibnamefont{Sasaki}}, 
\bibinfo{author}{\bibfnamefont{Y.}~\bibnamefont{Sakai}}, 
\bibinfo{author}{\bibfnamefont{H.}~\bibnamefont{Kouno}}, 
\bibnamefont{and}
\bibinfo{author}{\bibfnamefont{M.}~\bibnamefont{Yahiro}}, 
\bibinfo{journal}{Phys. Rev. \ D} 
\textbf{\bibinfo{volume}{84}},
\bibinfo{pages}{091901} (\bibinfo{year}{2011}); 
%%%%%%%%%%%%%%%%%%%%%%%%%%%%%%%%%%%%%%%%%%%%%%%%%%
%
%%%%%%%%%%%%%%%%%%%%%%%%%%%%%%%%%%%%%%%%%%%%%%%%%%
% [20] 
%%%%%%%%%%%%%%%%%%%%%%%%%%%%%%%%%%%%%%%%%%%%%%%%%%
\bibitem[{\citenamefont{Buisseret et al.}(2004)}]{BL}
\bibinfo{author}{\bibfnamefont{F.}~\bibnamefont{Buisseret}}, 
\bibnamefont{and}
\bibinfo{author}{\bibfnamefont{G.}~\bibnamefont{Lacroix}}, 
\bibinfo{journal}{Phys.\ Rev.\  D} \textbf{\bibinfo{volume}{85}},
\bibinfo{pages}{016009} (\bibinfo{year}{2012}).
%%%%%%%%%%%%%%%%%%%%%%%%%%%%%%%%%%%%%%%%%%%%%%%%%%
%
%%%%%%%%%%%%%%%%%%%%%%%%%%%%%%%%%%%%%%%%%%%%%%%%%%
% [21] 
%%%%%%%%%%%%%%%%%%%%%%%%%%%%%%%%%%%%%%%%%%%%%%%%%%
\bibitem[{\citenamefont{Sakai}(2010)}]{Sakai_hadron}
\bibinfo{author}{\bibfnamefont{Y.}~\bibnamefont{Sakai}},
\bibinfo{author}{\bibfnamefont{T.}~\bibnamefont{Sasaki}}, 
\bibinfo{author}{\bibfnamefont{H.}~\bibnamefont{Kouno}},
\bibnamefont{and}
\bibinfo{author}{\bibfnamefont{M.}~\bibnamefont{Yahiro}}, 
\bibinfo{journal}{J. Phys. \ G} 
\textbf{\bibinfo{volume}{39}},
\bibinfo{pages}{035004} (\bibinfo{year}{2012}).
%%%%%%%%%%%%%%%%%%%%%%%%%%%%%%%%%%%%%%%%%%%%%%%%%%
%
%%%%%%%%%%%%%%%%%%%%%%%%%%%%%%%%%%%%%%%%%%%%
% [22] 
%%%%%%%%%%%%%%%%%%%%%%%%%%%%%%%%%%%%%%%%%%%%
\bibitem[{\citenamefont{McLerran}(2007)}]{McLerran1}
\bibinfo{author}{\bibfnamefont{L.}~\bibnamefont{McLerran}},
\bibnamefont{and}
\bibinfo{author}{\bibfnamefont{R.}~\bibfnamefont{D.}~\bibnamefont{Pisarski}},
\bibinfo{journal}{Nucl. Phys.} \textbf{\bibinfo{volume}{A796}},
\bibinfo{pages}{83} (\bibinfo{year}{2007}); 
\bibinfo{author}{\bibfnamefont{Y.}~\bibnamefont{Hidaka}},
\bibinfo{author}{\bibfnamefont{L.}~\bibnamefont{McLerran}},
\bibnamefont{and}
\bibinfo{author}{\bibfnamefont{R.}~\bibfnamefont{D.}~\bibnamefont{Pisarski}},
\bibinfo{journal}{Nucl. Phys.} \textbf{\bibinfo{volume}{A808}},
\bibinfo{pages}{117} (\bibinfo{year}{2008}). 
%%%%%%%%%%%%%%%%%%%%%%%%%%%%%%%%%%%%%%%%%%%%
%
%%%%%%%%%%%%%%%%%%%%%%%%%%%%%%%%%%%%%%%%%%%%%%%%%%
% [23] 
%%%%%%%%%%%%%%%%%%%%%%%%%%%%%%%%%%%%%%%%%%%%%%%%%%
\bibitem{Nakano}
\bibinfo{author}{\bibfnamefont{E.}~\bibnamefont{Nakano}},
\bibnamefont{and}
\bibinfo{author}{\bibfnamefont{T.}~\bibnamefont{Tatsumi}}
  \bibinfo{journal}{Phys.\ Rev.\  D} \textbf{\bibinfo{volume}{71}},
  \bibinfo{pages}{114006} (\bibinfo{year}{2005}). 
%%%%%%%%%%%%%%%%%%%%%%%%%%%%%%%%%%%%%%%%%%%%%%%%%%%%%%%%%%%%%%%%%%%%%%%%%%%%%%%%%%%%%%%%%%%%%%%%%%%
%
%%%%%%%%%%%%%%%%%%%%%%%%%%%%%%%%%%%%%%%%%%%%%%%%%%
% [24] 
%%%%%%%%%%%%%%%%%%%%%%%%%%%%%%%%%%%%%%%%%%%%%%%%%%
\bibitem{Nickel}
\bibinfo{author}{\bibfnamefont{D.}~\bibnamefont{Nickel}},
  \bibinfo{journal}{Phys.\ Rev.\  Lett.} \textbf{\bibinfo{volume}{103}},
  \bibinfo{pages}{072301} (\bibinfo{year}{2009}),
  \bibinfo{journal}{Phys.\ Rev.\  D} \textbf{\bibinfo{volume}{80}},
  \bibinfo{pages}{074025} (\bibinfo{year}{2009}),
\bibinfo{author}{\bibfnamefont{S.}~\bibnamefont{Carignano}},
\bibinfo{author}{\bibfnamefont{D.}~\bibnamefont{Nickel}},
\bibnamefont{and}
\bibinfo{author}{\bibfnamefont{M.}~\bibnamefont{Buballa}}, 
  \bibinfo{journal}{Phys.\ Rev.\  D} \textbf{\bibinfo{volume}{82}},
  \bibinfo{pages}{054009} (\bibinfo{year}{2010}).   
%%%%%%%%%%%%%%%%%%%%%%%%%%%%%%%%%%%%%%%%%%%%%%%%%%
%
%%%%%%%%%%%%%%%%%%%%%%%%%%%%%%%%%%%%%%%%%%%%%%%%%%
% [25] 
%%%%%%%%%%%%%%%%%%%%%%%%%%%%%%%%%%%%%%%%%%%%%%%%%%
\bibitem[{\citenamefont{Kobayashi and Maskawa}(1970)}]{KMK}
\bibinfo{author}{\bibfnamefont{M.}~\bibnamefont{Kobayashi}}, 
\bibnamefont{and}
\bibinfo{author}{\bibfnamefont{T.}~\bibnamefont{Maskawa}},
  \bibinfo{journal}{Prog. Theor. Phys. } \textbf{\bibinfo{volume}{44}},
  \bibinfo{pages}{1422} (\bibinfo{year}{1970});
\bibinfo{author}{\bibfnamefont{M.}~\bibnamefont{Kobayashi}}, 
\bibinfo{author}{\bibfnamefont{H.}~\bibnamefont{Kondo}},
\bibnamefont{and}
\bibinfo{author}{\bibfnamefont{T.}~\bibnamefont{Maskawa}},
  \bibinfo{journal}{Prog. Theor. Phys. } \textbf{\bibinfo{volume}{45}},
  \bibinfo{pages}{1955} (\bibinfo{year}{1971}). 
%%%%%%%%%%%%%%%%%%%%%%%%%%%%%%%%%%%%%%%%%%%%%%%%%%
%
%%%%%%%%%%%%%%%%%%%%%%%%%%%%%%%%%%%%%%%%%%%%%%%%%%
% [26]
%%%%%%%%%%%%%%%%%%%%%%%%%%%%%%%%%%%%%%%%%%%%%%%%%%
\bibitem[{\citenamefont{'t Hooft}(1976)}]{tHooft}
\bibinfo{author}{\bibfnamefont{G.}~\bibnamefont{'t Hooft}},
  \bibinfo{journal}{Phys. Rev.\ Lett.} \textbf{\bibinfo{volume}{37}},
  \bibinfo{pages}{8} (\bibinfo{year}{1976});
  \bibinfo{journal}{Phys. Rev.\ D} \textbf{\bibinfo{volume}{14}},
  \bibinfo{pages}{3432} (\bibinfo{year}{1976});
  \textbf{\bibinfo{volume}{18}},
  \bibinfo{pages}{2199(E)} (\bibinfo{year}{1978}).
%%%%%%%%%%%%%%%%%%%%%%%%%%%%%%%%%%%%%%%%%%%%%%%%%%
%
%%%%%%%%%%%%%%%%%%%%%%%%%%%%%%%%%%%%%%%%%%%%%%%%%%%%%
% [27]
%%%%%%%%%%%%%%%%%%%%%%%%%%%%%%%%%%%%%%%%%%%%%%%%%%%%%
\bibitem[{\citenamefont{Boyd et al.}(1996)}]{Boyd}
\bibinfo{author}{\bibfnamefont{G.}~\bibnamefont{Boyd}},
\bibinfo{author}{\bibfnamefont{J.}~\bibnamefont{Engels}},
\bibinfo{author}{\bibfnamefont{F.}~\bibnamefont{Karsch}},
\bibinfo{author}{\bibfnamefont{E.}~\bibnamefont{Laermann}},
\bibinfo{author}{\bibfnamefont{C.}~\bibnamefont{Legeland}},
\bibinfo{author}{\bibfnamefont{M.}~\bibnamefont{L\"{u}tgemeier}},
\bibnamefont{and}
\bibinfo{author}{\bibfnamefont{B.}~\bibnamefont{Petersson}},
 \bibinfo{journal}{Nucl. Phys.} \textbf{\bibinfo{volume}{B469}},
\bibinfo{pages}{419} (\bibinfo{year}{1996}). 
%%%%%%%%%%%%%%%%%%%%%%%%%%%%%%%%%%%%%%%%%%%%%
%
%%%%%%%%%%%%%%%%%%%%%%%%%%%%%%%%%%%%%%%%%%%%%%%%%%%%%%%%%
% [28] 
%%%%%%%%%%%%%%%%%%%%%%%%%%%%%%%%%%%%%%%%%%%%%%%%%%%%%%%%%
\bibitem[{\citenamefont{Kaczmarek}(2002)}]{Kaczmarek}
\bibinfo{author}{\bibfnamefont{O.}~\bibnamefont{Kaczmarek}},
\bibinfo{author}{\bibfnamefont{F.}~\bibnamefont{Karsch}},
\bibinfo{author}{\bibfnamefont{P.}~\bibnamefont{Petreczky}},
\bibnamefont{and}
\bibinfo{author}{\bibfnamefont{F.}~\bibnamefont{Zantow}},  
  \bibinfo{journal}{Phys. Lett.\ B} \textbf{\bibinfo{volume}{543}},
  \bibinfo{pages}{41} (\bibinfo{year}{2002}).
%%%%%%%%%%%%%%%%%%%%%%%%%%%%%%%%%%%%%%%%%%%%%%%%%%%%%%%%%%%%%%%%%%
%
%%%%%%%%%%%%%%%%%%%%%%%%%%%%%%%%%%%%%%%%%%%%%%%%%%
% [29] 
%%%%%%%%%%%%%%%%%%%%%%%%%%%%%%%%%%%%%%%%%%%%%%%%%%
\bibitem[{\citenamefont{Borsanyi etal}(2010)}]{Borsanyi}
\bibinfo{author}{\bibfnamefont{S.}~\bibnamefont{Bors\'{a}nyi}}, 
\bibinfo{author}{\bibfnamefont{Z.}~\bibnamefont{Fodor}}, 
\bibinfo{author}{\bibfnamefont{C.}~\bibnamefont{Hoelbling}}, 
\bibinfo{author}{\bibfnamefont{S.}~\bibnamefont{D.}~\bibnamefont{Katz}}, 
\bibinfo{author}{\bibfnamefont{S.}~\bibnamefont{Krieg}}, 
\bibinfo{author}{\bibfnamefont{C.}~\bibnamefont{Ratti}}, 
\bibnamefont{and} 
\bibinfo{author}{\bibfnamefont{K.}~\bibnamefont{K.}~\bibnamefont{Szabo}},  
\bibinfo{howpublished}{arXiv:1005.3508 [hep-lat]} (\bibinfo{year}{2010}). 
%%%%%%%%%%%%%%%%%%%%%%%%%%%%%%%%%%%%%%%%%%%%%%%%%%
%
%%%%%%%%%%%%%%%%%%%%%%%%%%%%%%%%%%%%%%%%%%%%%%%%%%
% [30] 
%%%%%%%%%%%%%%%%%%%%%%%%%%%%%%%%%%%%%%%%%%%%%%%%%%
\bibitem[{\citenamefont{Soeldner}(2010)}]{Soeldner}
\bibinfo{author}{\bibfnamefont{W.}~\bibnamefont{S\"{o}ldner}}, 
\bibinfo{howpublished}{arXiv:1012.4484 [hep-lat]} (\bibinfo{year}{2010}). 
%%%%%%%%%%%%%%%%%%%%%%%%%%%%%%%%%%%%%%%%%%%%%%%%%%
%
%%%%%%%%%%%%%%%%%%%%%%%%%%%%%%%%%%%%%%%%%%%%%%%%%%
% [31]
%%%%%%%%%%%%%%%%%%%%%%%%%%%%%%%%%%%%%%%%%%%%%%%%%%
\bibitem[{\citenamefont{Kanaya}(2010)}]{Kanaya}
\bibinfo{author}{\bibfnamefont{K.}~\bibnamefont{Kanaya}}, 
\bibinfo{howpublished}{arXiv:hep-ph/1012.4235 [hep-ph]} (\bibinfo{year}{2010});
\bibinfo{howpublished}{arXiv:hep-ph/1012.4247 [hep-lat]} (\bibinfo{year}{2010}).
%%%%%%%%%%%%%%%%%%%%%%%%%%%%%%%%%%%%%%%%%%%%%%%%%%
%
%%%%%%%%%%%%%%%%%%%%%%%%%%%%%%%%%%%%%%%%%%%%%%%%%%
% [32] 
%%%%%%%%%%%%%%%%%%%%%%%%%%%%%%%%%%%%%%%%%%%%%%%%%%
\bibitem{Rehberg}
\bibinfo{author}{\bibfnamefont{P.}~\bibnamefont{Rehberg}},
\bibinfo{author}{\bibfnamefont{S.P.}~\bibnamefont{Klevansky}}
\bibnamefont{and}
\bibinfo{author}{\bibfnamefont{J.}~\bibnamefont{H\"{u}fner}}, 
  \bibinfo{journal}{Phys.\ Rev.\  C} \textbf{\bibinfo{volume}{53}},
  \bibinfo{pages}{410} (\bibinfo{year}{1996}). 
%%%%%%%%%%%%%%%%%%%%%%%%%%%%%%%%%%%%%%%%%%%%%%%%%%%%%%%%%%%%%%%%%%%%%%%%%%%%%%%%%%%%%%%%%%%%%%%%%%%


\end{thebibliography}
\end{document}